\documentclass[12pt,preprint]{aastex}

\received{2000 December 20}
\accepted{2001 ??}

\shorttitle{Quasi-self-similar evolution of the correlation function}
\shortauthors{Suginohara et al.}

\begin{document}
\title{Quasi-Self-Similar Evolution of the Two-Point Correlation
Function:\\ Strongly Nonlinear Regime in $\Omega_0 < 1$ Universes}
\author{Tatsushi Suginohara\altaffilmark{1},
Atsushi Taruya, and Yasushi Suto}
\affil{Department of Physics and Research Center for the Early Universe, 
School of Science, University of Tokyo, Tokyo 113-0033, Japan.}
\email{tatsushi@res.otaru-uc.ac.jp, ataruya@utap.phys.s.u-tokyo.ac.jp,
       suto@phys.s.u-tokyo.ac.jp}
\altaffiltext{1}{Present address: 
Faculty of Commerce, Otaru University of Commerce, Otaru 047-8501, Japan.}
\begin{abstract}
  The well-known self-similar solution for the two-point correlation
  function of the density field is valid only in an Einstein -- de
  Sitter universe.  We attempt to extend the solution for non --
  Einstein -- de Sitter universes.  For this purpose we introduce an
  idea of quasi-self-similar evolution; this approach is based on the
  assumption that the evolution of the two-point correlation is a
  succession of stages of evolution, each of which spans a short enough
  period to be considered approximately self-similar.  In addition we
  assume that clustering is stable on scales where a physically
  motivated  `virialization condition' is satisfied.  These
  assumptions lead to a definite prediction for the behavior of the
  two-point correlation function in the strongly nonlinear regime.  We
  show that the prediction agrees well with N-body simulations in non --
  Einstein -- de Sitter cases, and discuss some remaining problems.
\end{abstract}
\keywords{cosmology: theory --- large-scale structure of universe 
 --- gravitation --- methods: N-body simulations}


%
%
\clearpage
\section{Introduction}

It is generally believed that the structure in our universe has grown
out of tiny density fluctuations through gravitational instability.  The
growth of those density fluctuations is completely described by linear
theory on scales much larger than the correlation length of the density
field (e.g., Peebles 1980).  The behavior of the nonlinear density field
on smaller scales, however, needs to be modeled with additional
assumptions, and the resulting predictions should be verified and
calibrated through extensive comparison with N-body simulations.

The most popular prediction in the nonlinear regime is based on the
combination of the self-similar evolution and the stable clustering
ansatz \citep{pee74,pee80,dav77}; if the linear power spectrum of
density field follows a single power law $\propto k^n$, and the universe
is described by the Einstein -- de Sitter model, the evolution proceeds
in a self-similar manner since there is no characteristic scale in the
system.  This self-similarity, together with the ansatz that the
clustering is stable in the strongly nonlinear regime, predicts that the
logarithmic slope of the two-point correlation function $\xi$ is equal
to $-3(n+3)/(n+5)$.

This solution has been widely applied in modeling the nonlinear
gravitational clustering \citep{ham91,pea94,pea96,jai95} and in
understanding the pair-wise velocity dispersions and thus the
redshift-space distortion \citep{sut97,jin98b}. In fact, the above
prescription has been applied even in the cases where the universe is
not described by the Einstein -- de Sitter model, and/or the linear
power spectrum is not of a power-law form.  The fitting formula for the
nonlinear power spectrum by \citet{pea96}, for instance, takes account
of the non power-law nature of the linear spectrum, but they have not
examined the validity of the stable clustering solution in non Einstein
-- de Sitter models because `if collapse occurs at high redshift, then
$\Omega=1$ may be assumed at that time' \citep{pea94}.

We revisit this issue in detail for the non -- Einstein -- de Sitter
case, i.e., when the matter density parameter $\Omega_0$ is smaller than
unity.  Using N-body simulations with $64^3$ particles, \citet{sug91}
already found that while the stable solution is reproduced well in the
Einstein de -- Sitter model, the slope of $\xi$ becomes steeper in
$\Omega_0 < 1$ models (see also Suto 1993).  This is not surprising at
all because the derivation of the solution heavily relies on the
scale-free nature of the Einstein -- de Sitter model, that is, (1) $a
\propto t^{2/3}$, where $a$ is the cosmic scale factor, and (2) $D(t)
\propto a$, where $D(t)$ is the linear growth rate of density
fluctuations.  This motivates us to find a suitable modification of the
self-similar solution which is also applicable to the non -- Einstein --
de Sitter case.

In this paper we introduce an idea of quasi-self-similarity.  This is
based on the assumption that the evolution of the two-point correlation
proceeds as a sequence of different quasi-self-similar stages, each of
which is described by the {\it locally} self-similar solution determined
by the cosmological parameters at that epoch.  We also assume that the
clustering becomes stable on scales where a `virialization condition' is
satisfied.  These two assumptions lead to a prediction of the behavior
of the two-point correlation function in the strongly nonlinear regime.
We show the extent to which the prediction agrees with the results of
high-resolution N-body simulations.

%
%
\section{Quasi-Self-Similar Evolution}

\subsection{The Self-Similar Solution in the Einstein--de Sitter Model}

For later convenience, we first briefly review the self-similar solution
in the Einstein-- de Sitter case \citep{pee74,pee80,dav77}.  Suppose a
system of particles with mass $m$.  In the fluid limit, the system is
described by the one-particle distribution function $f({\bf x},{\bf
p},t)$, where ${\bf x}$ is the comoving coordinate, and ${\bf p} \equiv
ma^2d{\bf x}/dt$ is its canonical momentum.  The distribution function
$f({\bf x},{\bf p},t)$ obeys the Vlasov equation \citep{ina76,pee80}:
\begin{eqnarray}
\label{eqn:vlasov}
\frac{\partial f}{\partial t}
  + \frac{p_i}{ma^2} \frac{\partial f}{\partial x_i}
  - m \frac{\partial \phi}{\partial x_i}
      \frac{\partial f}{\partial p_i} = 0 ,
\end{eqnarray}
where the gravitational potential $\phi$ satisfies
\begin{eqnarray}
\nabla^2 \phi = 4\pi G m a^{-1}\int f\,d^3 p .
\end{eqnarray}

In the Einstein -- de Sitter model ($a \propto t^{2/3}$), equation
(\ref{eqn:vlasov}) admits a self-similar solution of the form:
\begin{eqnarray}
\label{eq:ssf}
f({\bf x},{\bf p},t) 
  = t^{-3\alpha-1} \hat f ({\bf x}/t^{\alpha}, {\bf p}/t^{\alpha+1/3}) .
\end{eqnarray}
The value of the parameter $\alpha$ is determined by matching the
solution in a linear regime as follows; first note that equation
(\ref{eq:ssf}) implies the following form for the two-point correlation
function $\xi (x,t)$:
\begin{eqnarray}
\label{eqn:xi}
\xi (x,t) = \hat \xi (x/t^\alpha) .
\end{eqnarray}
If the initial power spectrum of density fluctuations, $P_{\rm
  initial}(k)$, is proportional to $k^n$, then $\xi(x,t)$ at later
  epochs behaves as $x^{-(n+3)} t^{4/3}$ on large scales where linear
  theory is valid.  Thus the value of $\alpha$ is explicitly specified as
\begin{eqnarray}
\label{eqn:alpha}
\alpha = {4 \over 3(n+3)} .
\end{eqnarray}

In the small scale (nonlinear) limit, it is often assumed that the
average proper separation of pairs remains constant (the stable
clustering hypothesis):
\begin{eqnarray}
\label{eqn:stable}
\langle v_{21}(x,t) \rangle = - \dot a x ,
\end{eqnarray}
where $v_{21}$ is the relative peculiar velocity for a pair, and
the brackets denote the average over pairs at a given comoving
separation $x$.  This assumption with equation (\ref{eqn:xi}) fixes the
behavior of the correlation function in the nonlinear limit.
Substituting equation (\ref{eqn:stable}) into the equation of the
particle pair conservation:
\begin{eqnarray}
{\partial \xi \over \partial t}
  + {1 \over a x^2} 
    {\partial \over \partial x}
    [x^2 (1+\xi) \langle v_{21}(x,t) \rangle] = 0,
\end{eqnarray}
yields
\begin{eqnarray}
{\partial \xi \over \partial t}
= {\dot a \over a}{ 1\over  x^2} 
   {\partial \over \partial x} [x^3 \xi ],
\end{eqnarray}
for $\xi \gg 1$.  Thus $\xi$ in the nonlinear regime should be of the
form
\begin{eqnarray}
\label{eqn:a3g}
\xi = a^3 g(ax), 
\end{eqnarray}
where $g$ is an arbitrary function at this point. Finally the
consistency with the above similarity solution requires that $\xi$
should be given by the following power-law form:
\begin{eqnarray}
\label{eqn:similarity}
\xi (x,t) \propto x^{- \frac{3(n+3)}{n+5}}\, t^{\frac 4{n+5}}
          \propto x^{- \frac{3(n+3)}{n+5}}\, a^{\frac 6{n+5}} .
\end{eqnarray}
\subsection{Quasi-Self-Similar Clustering Solutions}

Now we attempt to generalize the above self-similar solution in a
hypothetical universe where $a \propto t^p$ and $D(t) \propto t^q$.  In
non--Einstein--de Sitter models, both $p$ and $q$ are not constant but
change with time.  As long as the time-scale of the change of $p$ and
$q$ is smaller than that of the cosmic expansion, however, they can be
regarded as constants for a short period at each epoch.

We begin with the assumption of similarity for $\xi(x,t)$:
\begin{eqnarray}
\label{eqn:similarity_xi}
\xi (x,t) = \hat \xi (x/t^{\alpha}).
\end{eqnarray}
Unlike in the Einstein -- de Sitter case, any kind of self-similarity
does not hold in a strict sense.  However, the range of $p$ and $q$
considered here is close to that in the Einstein -- de Sitter case, so
it seems reasonable to assume that some form of self-similarity is
realized in an approximate sense.  The simplest possibility is that,
as in the previous subsection, $\xi(x,t)$ has the form
(\ref{eqn:similarity_xi}).  With this ansatz we repeat the same procedure
in the previous subsection, adopting $P_{\rm initial}(k) \propto k^n$
and the stable clustering hypothesis in the strongly nonlinear regime.

In this case, one has $\xi \propto x^{-(n+3)} t^{2q}$ in linear regime,
and equation (\ref{eqn:alpha}) is replaced by
\begin{eqnarray}
\alpha = {2q \over n+3} .
\end{eqnarray}
Combining the form (\ref{eqn:a3g}) for $\xi$ in the strongly nonlinear
regime yields, instead of equation (\ref{eqn:similarity}),
\begin{eqnarray}
\label{eqn:qsspl}
\xi (x,t) \propto x^{- \frac{3(n+3)}{n+3 + 2f}} \, a^{\frac{6f}{n+3 + 2f}} 
= (ax)^{- \frac{3(n+3)}{n+3 + 2f}}\, a^3 ,
\end{eqnarray}
for $\xi \gg 1$.  The quantity $f\equiv q/p$ in the above expression is
in fact the familiar logarithmic derivative of $D$ with respect to $a$.
An excellent approximation to $f$ is \citep{lah91}
\begin{eqnarray}
\label{eqn:fomega}
f = \frac{d \ln D}{d \ln a} \sim \Omega(a)^{0.6} 
+ {\lambda(a) \over 70} \left[1+\frac{\Omega(a)}{2}\right] ,
\end{eqnarray}
where
\begin{eqnarray}
\Omega(a) = \frac{\Omega_0}{\Omega_0 
+ (1 - \Omega_0 - \lambda_0)(a/a_0) + \lambda_0(a/a_0)^3}
\end{eqnarray}
and
\begin{eqnarray}
\lambda(a) = \frac{\lambda_0(a/a_0)^3}{\Omega_0 
+ (1 - \Omega_0 - \lambda_0)(a/a_0) + \lambda_0(a/a_0)^3} ,
\end{eqnarray}
with $\lambda_0$ being the dimensionless cosmological constant at the
present epoch $a_0$.

\subsection{Comparison with N-body Simulations}
\label{subsect:n-body}

Let us compare in detail the self-similar solution
(eq.~[\ref{eqn:similarity}]) and the quasi-self-similar solution
(eq.~[\ref{eqn:qsspl}]) with high-resolution N-body simulations (Jing
1998; 2001b in preparation).  The simulations consist of two Einstein --
de Sitter models, two open models ($\Omega_0 = 0.1$, $\lambda_0 = 0$),
and two spatially flat models ($\Omega_0 = 0.1$, $\lambda_0 = 0.9$).
All the models employ scale-free initial power spectra $P_{\rm
initial}(k) \propto k^n$ with $n=-1$ and $n=-2$.  Gravitational force
calculation is based on the ${\rm P^3M}$ algorithm.  Simulation
parameters are listed in Table~\ref{tab:param}, and further details of
the simulations are described in Jing (1998, 2001b in preparation).

\begin{figure}
\plotone{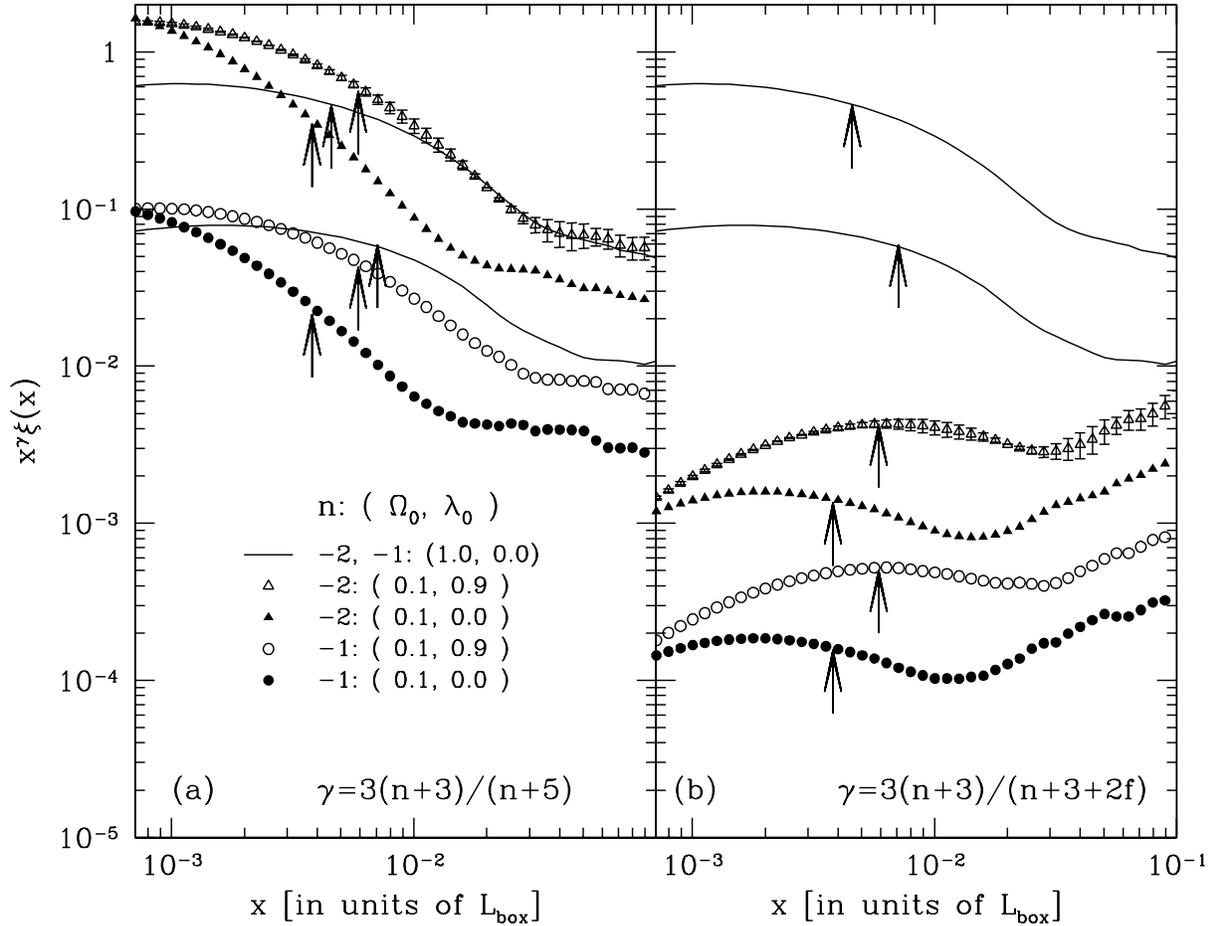}
\figcaption{Two-point correlation functions in the simulations
  scaled according to
  (a) conventional self-similar solution,
  and (b) quasi-self-similar solution with scale independent
  slope.
  The solid lines are the results for $\Omega_0 = 1$ models, and
  the symbols are those for $\Omega_0 = 0.1$ models with
  either $\lambda_0 = 0$ or $\lambda_0 = 0.9$.
  Error bars are estimated from three realizations for each model;
  only the error bars for $n=-2$ and $(\Omega_0,\lambda_0) = (0.1,
  0.9)$ are shown.  Other models have smaller error bars, which are
  not shown for clarity.
  The arrows correspond to $\xi(x)=100$ for each model. 
\label{fig:xisc}}
\end{figure}

Figure~\ref{fig:xisc}a plots the two-point correlation functions which
are appropriately scaled with respect to $x$ according to the
conventional self-similar solution (eq.[\ref{eqn:similarity}]).  It is
clear that while $\Omega_0 = 1$ simulations are in good agreement with
equation~(\ref{eqn:similarity}) beyond $\xi = 100$ (indicated by an
arrow for each model), $\Omega_0 = 0.1$ models are not; the disagreement
is particularly severe for the $\lambda_0 = 0$ models.
Figure~\ref{fig:xisc}b shows the similar plot but with scaling for
quasi-self-similar solution (eq.[\ref{eqn:qsspl}]), where we use the
value of $f$ evaluated at the present epoch.  The figure indicates that
the agreement with the simulations in $\Omega_0 = 0.1$ models indeed
improves.  Nevertheless the results in $\Omega_0 = 0.1$ and $\lambda_0 =
0.9$ models do not yet show acceptable agreement with
equation~(\ref{eqn:qsspl}). The origin of the disagreement is studied by
Suto, Taruya, \& Suginohara (2001).

This comparison points to the following two suggestions; (i) in
$\Omega_0 < 1$ models the conventional self-similar solution fails to
describe the behavior of $\xi$ for $\xi \gtrsim 100$, and (ii) taking
account of the dependence of the slope on the cosmological parameters
as in equation~(\ref{eqn:qsspl}) does not yet yield an
acceptable prediction.  
With these points in mind we attempt to improve the
quasi-self-similar solution in the next section.

%
%
\section{An Improved Model}
\label{sect:qss}

\subsection{Virialization Condition and
Scale Dependence of the Slope of $\xi$}
\label{subsect:scaledep}

In the previous section we have applied the quasi-self-similarity only
at the present epoch. In reality, however, the logarithmic slope of
the correlation function:
\begin{eqnarray}
\label{eqn:dxi}
\frac{d\ln \xi}{d\ln x} = - \frac{3(n+3)}{n+3 + 2f(a_{\rm vir}(x))}
\end{eqnarray}
should be fixed locally at the epoch of the virialization of the
corresponding scale, $a_{\rm vir}(x)$.  This naturally generates
additional scale-dependence on the resulting solution through the
time-dependence of $f$.  More specifically, we attempt to incorporate
this effect and to improve the model in the previous subsection as
follows.

In order to determine $a_{\rm vir}(x)$ at each scale $x$, we need to
make an assumption about the scale on which the system has been
virialized at a given epoch $a$.  Here we mainly adopt the following
assumption: the system has just been virialized at a scale 
$x_{\rm vir}(a)$ where the volume average of the two-point correlation
function reaches the critical overdensity of a virialized halo,
$\Delta_{\rm vir}(a)$, predicted by the spherical collapse model:
\begin{eqnarray}
\label{eqn:virial}
\bar{\xi}(x_{\rm vir}(a))  = \Delta_{\rm vir}(a) ,
\end{eqnarray}
where
\begin{eqnarray}
\bar{\xi}(x) \equiv \frac{3}{x^3} 
\int_0^x y^2 \xi(y) d y .
\end{eqnarray}
It seems natural to use $\bar{\xi}$ rather than $\xi$ for the present
purpose because the right-hand-side of equation (\ref{eqn:virial}) is
the {\it average}\/ density of a halo that has just been virialized.  In
$\lambda_0 = 0$ models, the critical density $\Delta_{\rm vir}(a)$ is
explicitly written as \citep{lac93}
\begin{eqnarray}
\Delta_{\rm vir}(a) & = & 4\pi^2 \frac{(\cosh \eta - 1)^3}
{(\sinh \eta - \eta)^2} , \\
\cosh \eta & = & \frac{2}{\Omega(a)} - 1 ,
\end{eqnarray}
and its accurate fitting formula in $\lambda_0 = 1 - \Omega_0$ models
\citep{nak97} is given by
\begin{eqnarray}
\Delta_{\rm vir}(a) \simeq 18\pi^2 
\Biggl\lbrace 1 + 0.4093 \biggl[\frac{1}{\Omega(a)}-1\biggr]^{0.9052}
\Biggr\rbrace .
\end{eqnarray}
The `virialization condition' (\ref{eqn:virial}) may be 
somewhat arbitrary, but is perhaps most 
natural and physically motivated among other choices.

Furthermore we assume that, for $x \leq x_{\rm vir}(a)$, the stable
clustering
condition is satisfied, i.e., $\xi(x,a)$ is of the form (\ref{eqn:a3g}).
This is equivalent to saying that, at a fixed physical scale $r=ax$, the
slope of $\xi$ is kept constant while its amplitude grows in proportion
to $a^3$ (c.f., eq.[\ref{eqn:qsspl}]).  

The above assumptions are not yet sufficient in predicting $\xi(x)$ for
a given model.  The amplitude of $\xi$ at an arbitrary point needs to be
specified by hand because our model does not predict the overall
amplitude of $\xi(x)$.  Once the value of $\xi$ is fixed at a scale
$x=\varepsilon$ (for example, in the next subsection, we take
$\varepsilon_{\rm grav}$, the gravitational softening length, as
$\varepsilon$ and use the value of $\xi(\varepsilon_{\rm grav})$ in the
simulations), we can compute $\xi(x)$ at $a = a_0$ up to $x_{\rm vir,0}
\equiv x_{\rm vir}(a_0)$.  The procedure is as follows:

1. {\it Set boundary condition}. --- Below an extremely small scale $x_l$,
where $\xi$ is sufficiently large, our model prediction should reduce
to the conventional self-similar solution in the Einstein -- de Sitter
universe; i.e., 
\begin{eqnarray}
\label{eqn:boundary}
\xi \propto x^{-3(n+3)/(n+5)} \qquad \mbox{for}\  x \le x_l .
\end{eqnarray}
This is because the epoch of the virialization of the corresponding
scale is sufficiently early, the universe is indistinguishable from the
Einstein -- de Sitter model \citep{pea94}.  Equation
(\ref{eqn:boundary}) implies that
\begin{eqnarray}
\bar\xi(x_l) = \frac{(n+5)}{2}\xi(x_l) .
\end{eqnarray}
We first choose $x_l$ arbitrarily and set the value of $\xi(x_l)$ to
be sufficiently large, say, $10^8$.  This is the starting point of our
computation.

2. {\it Compute $a_{\rm vir}(x)$}. --- From the above assumptions it
follows that
\begin{eqnarray}
\label{eqn:qss2}
\bar{\xi}(x) & = & \biggl( \frac{a_{\rm vir}(x)}{a_0} \biggr)^{-3}
    \Delta_{\rm vir}(a_{\rm vir}(x)) .
\end{eqnarray}
At the current scale $x$, we solve equation (\ref{eqn:qss2})
for $a_{\rm vir}(x)$.

3. {\it Advance $x$} . --- Substituting the value of $a_{\rm vir}(x)$ 
into equation (\ref{eqn:dxi}) gives the local slope $d\ln\xi/d\ln x$ at $x$.
Then we can advance $x$ by a small interval $\Delta x$
and compute $\xi(x+\Delta x)$, and then $\bar\xi(x+\Delta x)$.

4. We repeat the procedures 2 and 3 until $\bar\xi(x)$ becomes
equal to $\Delta_{\rm vir}(a_0)$, which corresponds to
the present virialization scale $x_{\rm vir,0}$.

5. {\it Normalization}. --- Finally we shift the resulting solution
so that it can match the given amplitude at $x = \varepsilon$,
using the fact that, 
if $\xi^{(1)}(x)$ is a solution to equation (\ref{eqn:dxi}),
so is $\xi^{(2)}(x) \equiv
\xi^{(1)}(\alpha x)$ with an arbitrary constant $\alpha$.

\subsection{Comparison with N-body Simulations}
\label{subsect:qss_n-body}

\begin{figure}
\plotone{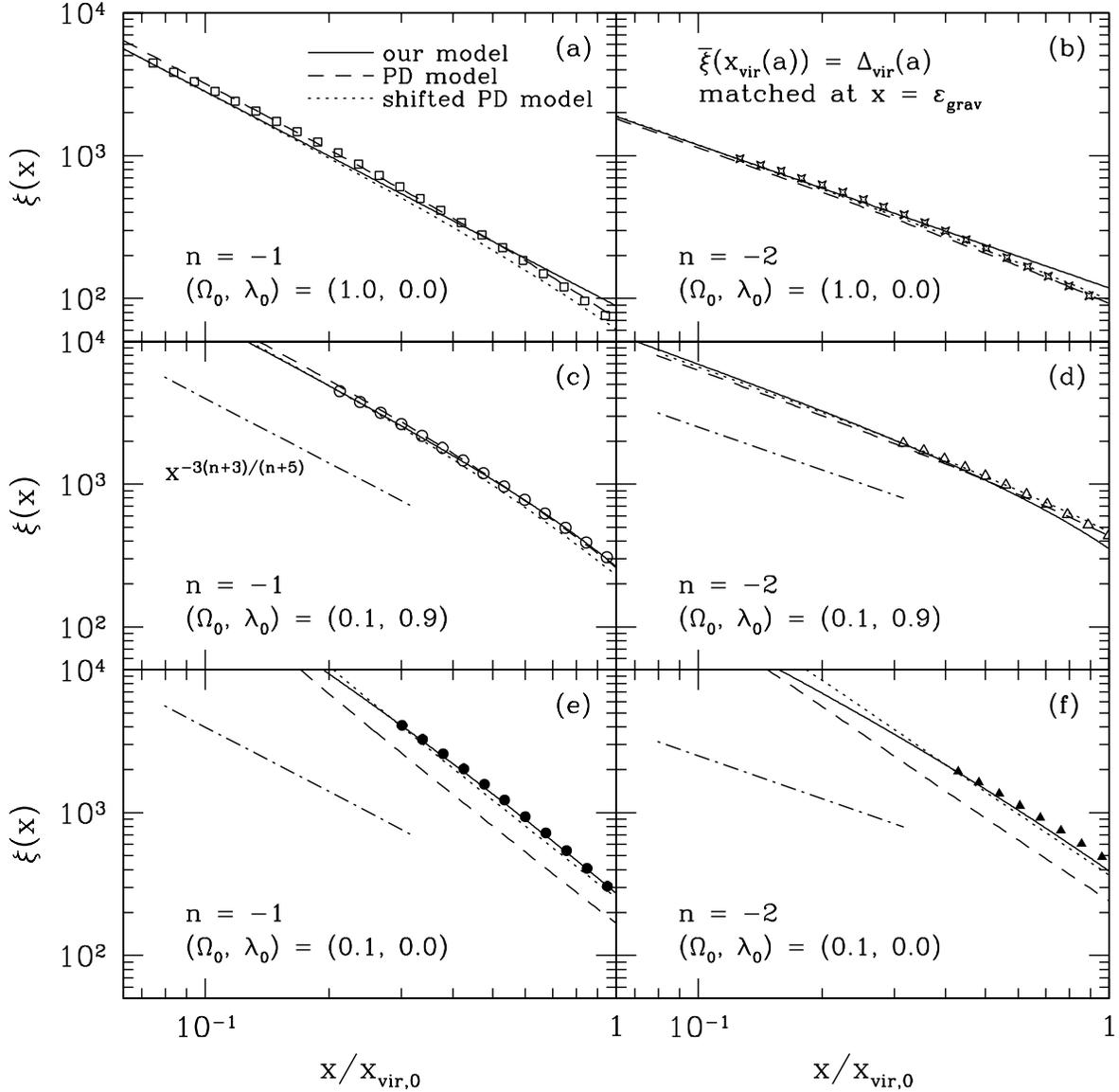}
\figcaption{Quasi-self-similar solutions with a virialization condition
  $\bar{\xi}(x_{\rm vir}(a)) = \Delta_{\rm vir}(a)$.
  The solid lines are obtained by numerically integrating 
  eq.~[\ref{eqn:dxi}].
  The length scale $x$ is normalized by $x_{\rm vir,0}$.
  The dashed lines are computed using the Peacock-Dodds formula.
  The dotted lines also show the Peacock-Dodds formula,
  but this time each result has been horizontally shifted to match 
  the simulation at $x = \varepsilon_{\rm grav}$.
  The dot-dashed lines show the slope of the conventional self-similar
  solution.
  The symbols are the results of the simulations.
  (a) $n=-1$ and $(\Omega_0,\lambda_0) = (1.0, 0.0)$.
  (b) $n=-2$ and $(\Omega_0,\lambda_0) = (1.0, 0.0)$.
  (c) $n=-1$ and $(\Omega_0,\lambda_0) = (0.1, 0.9)$.
  (d) $n=-2$ and $(\Omega_0,\lambda_0) = (0.1, 0.9)$.
  (e) $n=-1$ and $(\Omega_0,\lambda_0) = (0.1, 0.0)$.
  (f) $n=-2$ and $(\Omega_0,\lambda_0) = (0.1, 0.0)$.
\label{fig:xiqss1}}
\end{figure}

Figure~\ref{fig:xiqss1} compares our improved model predictions with the
N-body results.  Using the scaling relation described above, we match
the amplitude of our solution to that of the simulations at $x =
\varepsilon_{\rm grav}$ for each model (the innermost symbols).  The
length scale $x$ is normalized by $x_{\rm vir,0}\equiv x_{\rm vir}(a_0)$
and we show the results only in the virialized regime, $x<x_{\rm
vir,0}$.  Clearly, our predictions are in good agreement with
simulations for $\xi \gtrsim 200$ in all models. In particular the
predicted dependence on the spectral index $n$ is excellently reproduced
in the simulation results.  Given the simplicity of our procedure, this
may be regarded as a considerable success.  For $\xi < 200$, however,
the slopes of our predicted correlation functions are shallower than
those of simulations (Figs.~\ref{fig:xiqss1}a and~\ref{fig:xiqss1}b).

Let us compare our model predictions with the conventional self-similar
solution (eq.~[\ref{eqn:similarity}]) in more detail.  The power-law
slope of the conventional self-similar solution is shown by the
dot-dashed lines in Figure~\ref{fig:xiqss1}.  In $\Omega_0 = 0.1$ and
$\lambda_0 = 0.9$ cases both our model and the simulation have roughly
the same slope as the conventional one near $x = \varepsilon_{\rm
grav}$, but in both of them the slope becomes steeper as the scale
approaches to $x_{\rm vir,0}$.  More impressively, in $\Omega_0 = 0.1$
and $\lambda_0 = 0.0$ cases both our model and the simulation have a
steeper slope on all scales shown in the figure than the conventional
one.

Also shown in Figure~\ref{fig:xiqss1} are the correlation functions
obtained by Fourier-transforming the fitting formulae for the nonlinear
power spectra by \citet{pea96}.  The Peacock-Dodds formula is extremely
useful because it gives not only the shape but also the amplitude of the
two-point correlation function from linear to strongly nonlinear
regimes.  The agreement between the Peacock-Dodds formula and the
simulations is very well except for $\Omega_0 = 0.1$ and $\lambda_0 = 0$
models.  In these cases the Peacock-Dodds formula systematically
underpredicts the simulation results. Nevertheless this could be
adjusted somehow by shifting the amplitude of the Peacock-Dodds formula
so as to match the simulation at $x = \varepsilon_{\rm grav}$ (see the
dotted lines in Fig.~\ref{fig:xiqss1}).  Rather an important advantage
of our model over the Peacock-Dodds prescription is that it does
successfully {\it predict}\/ the slope of $\xi$ up to a scale where
deviation from the conventional one, $-3(n+3)/(n+5)$, is significant.
Note that the slope of $\xi$ computed from the Peacock-Dodds formula is,
on scales where it deviates from $-3(n+3)/(n+5)$, simply an
interpolation from the numerical simulations.  In this sense, our result
implies that there is room for further improvement in the original
Peacock-Dodds formula, and our present model may be useful for that
purpose.

\begin{figure}
\plotone{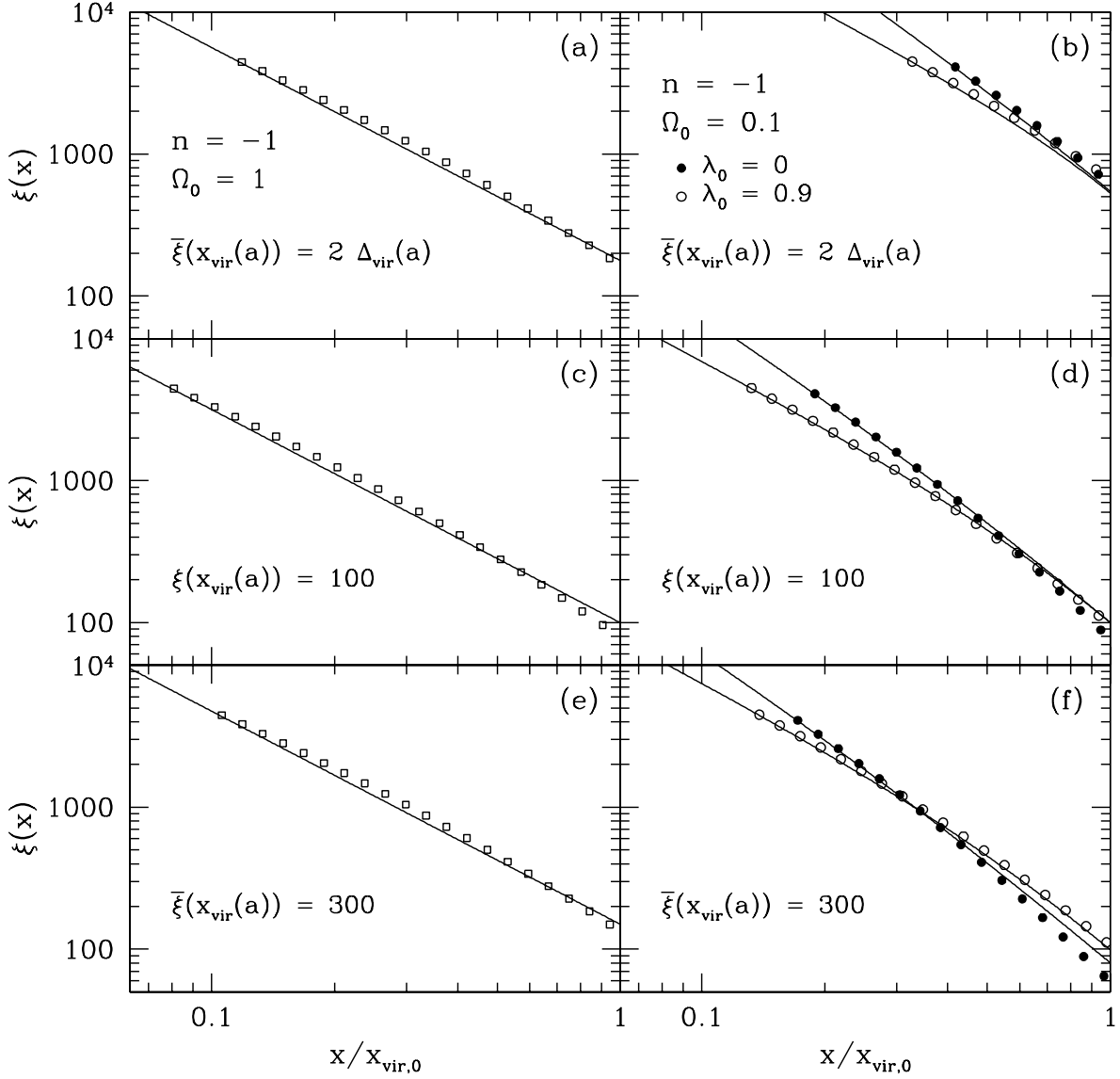}
\figcaption{Quasi-self-similar solutions for $n=-1$.
  The solid lines are obtained by numerically integrating 
  eq.~[\ref{eqn:dxi}].
  The symbols are the results of the simulations.
  (a) $\Omega_0 = 1$ model with a virialization condition
      $\bar{\xi}(x_{\rm vir}(a)) = 2 \Delta_{\rm vir}(a)$.
  (b) $\Omega_0 < 1$ models with 
      $\bar{\xi}(x_{\rm vir}(a)) = 2 \Delta_{\rm vir}(a)$.
  (c) $\Omega_0 = 1$ model with 
      $\xi(x_{\rm vir}(a)) = 100$.
  (d) $\Omega_0 < 1$ models with 
      $\xi(x_{\rm vir}(a)) = 100$.
  (e) $\Omega_0 = 1$ model with 
      $\bar{\xi}(x_{\rm vir}(a)) = 300$.
  (f) $\Omega_0 < 1$ models with 
      $\bar{\xi}(x_{\rm vir}(a)) = 300$.
\label{fig:xiqss2}}
\end{figure}

\begin{figure}
\plotone{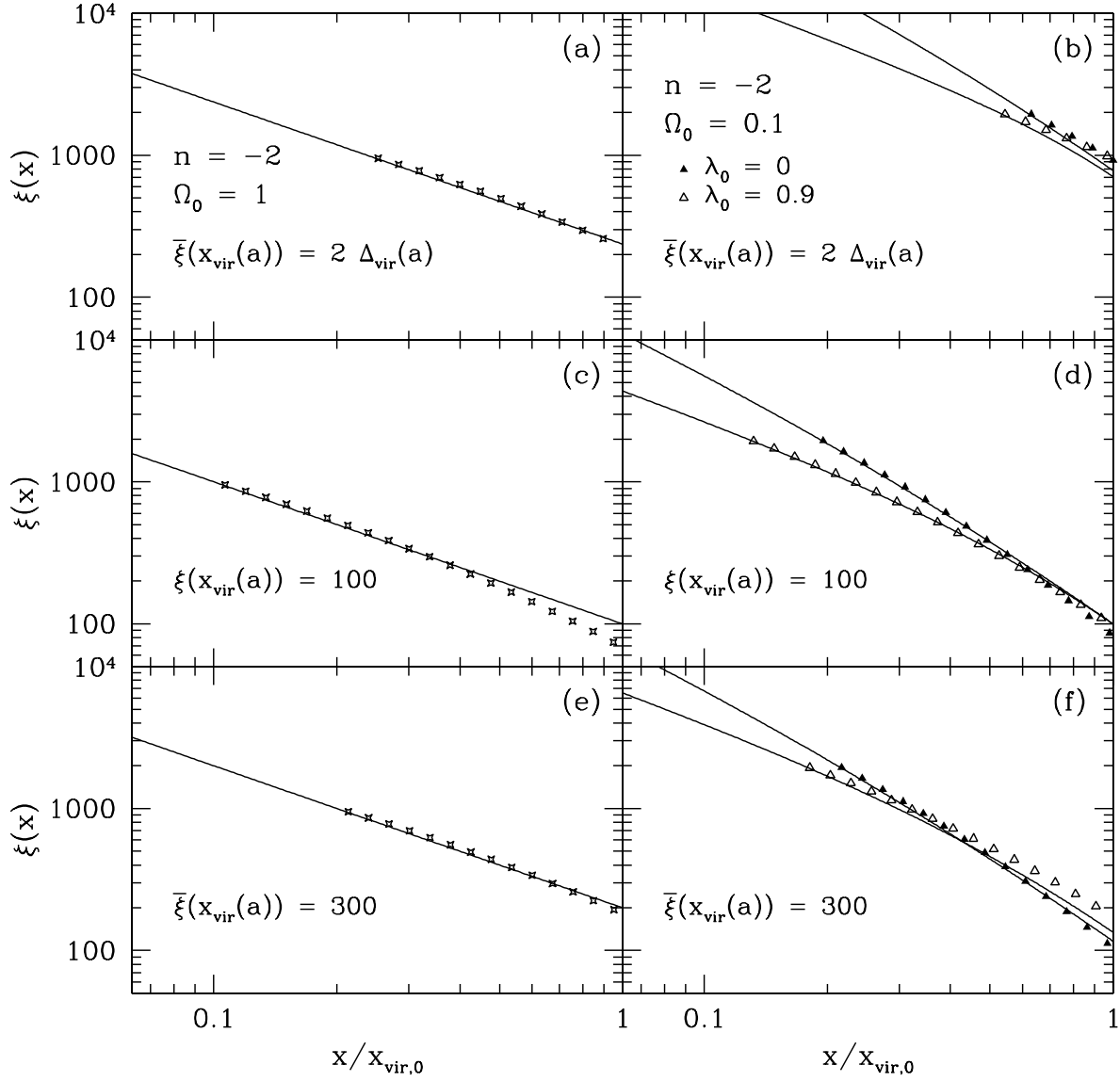}
\figcaption{As for Fig.~\ref{fig:xiqss2}, but for $n=-2$.
\label{fig:xiqss3}}
\end{figure}

One may wonder if it is possible to improve our model predictions by
varying the condition (\ref{eqn:virial}) somehow.
Figures~\ref{fig:xiqss2} and~\ref{fig:xiqss3} present those results.
The top panels in those figures adopt $\bar{\xi}(x_{\rm vir}(a)) = 2
\Delta_{\rm vir}(a)$ instead of equation (\ref{eqn:virial}).
While the agreement between the quasi-self-similar prediction and the
simulations is indeed improved for $\Omega_0=1$ models, this is not the
case for $\Omega_0 < 1$ models.

The middle and bottom panels show the results for 
$\xi(x_{\rm vir}(a)) = 100$, and
$\bar{\xi}(x_{\rm vir}(a)) = 300$, respectively, instead of equation
(\ref{eqn:virial}).  In each case we obtain acceptable agreement for
both $\Omega_0 = 1$ and $\Omega_0 < 1$ models simultaneously, although
the modified conditions lose the physical basis and should be regarded
as empirical at best.

%
%
\subsection{Validity of the stable clustering hypothesis}
\label{sect:stable}

The stable clustering hypothesis is an essential ingredient in our
model, but has somewhat been in doubt in the recent literature
\citep{yan00,maf00,cal01}.  We argue here, nevertheless, that the
hypothesis still remains to be a reasonable assumption.

\citet{yan00} relate the inner density profile of virialized halos with
the velocity parameter $h \equiv -\langle v_{21} \rangle / \dot a x$.
They claim that $h$ should approach zero in the nonlinear limit if
the logarithmic slope of the inner density profile is larger than
$-3/2$.  As pointed out by \citet{maf00}, their argument is valid only
when all halos have an equal mass, and should be completely altered for
a realistic mass function.

On the basis of simulations by \citet{jai97}, \citet{cal01} find for a
variety of cosmological models a universal relation between $f \bar \xi$
and $h$, and propose a fitting formula describing the relation.
Although extrapolating this formula implies that $h \rightarrow 0$ in
the nonlinear limit, they claim that their formula is valid for $f
\bar\xi \lesssim 10^3$.  Thus their results are not inconsistent with
the idea that $h \sim 1$ in the strongly nonlinear regime.  In fact,
more recent numerical work supports the stable clustering assumption
(Jing 2001a; Fukushige \& Suto 2001).  

%
%
\section{Discussion}
\label{sect:discussion}

We have shown that the conventional self-similar solution in the
Einstein -- de Sitter universe does not describe the behavior of
two-point correlation functions in $\Omega_0 \not= 1$ models for
strongly nonlinear regimes of cosmological interest, $\xi < 10^4$.
Instead, we have proposed a simple model to describe the two-point
correlation functions in strongly nonlinear regimes by introducing the
quasi-self-similar ansatz.  In fact we have shown that the resulting
model predictions in non -- Einstein -- de Sitter universes agree better
with the high-resolution N-body simulations.

On the other hand, our current model is not fully successful yet in the
sense that the predicted behavior for $\xi < 200$ systematically differs
from that observed in simulations. Empirically this situation can be
improved by an appropriate choice of the virialization threshold
$\bar\xi_{\rm vir}$. While the physical meaning of this procedure is not
clear, this may be related to some other physics that we omit in the
present simple prescription. After all the regime for $\xi < 200$ may
not be completely dominated by the stable clustering evolution, and the
effect of linear and quasi-linear evolution is likely to be important as
well. In fact it may be the case that the stable condition
(\ref{eqn:stable}) is not realized instantaneously when the virial
condition (\ref{eqn:virial}) is satisfied, and that the slope at the
corresponding scale approaches the value implied by equation
(\ref{eqn:dxi}) only gradually as the scale becomes more strongly
nonlinear (for instance, $\xi > 200$) and completely decoupled from the
linear evolution of the environment. We plan to check this
interpretation using the time-evolution of the simulation results in due
course.

One of the most important applications of the conventional self-similar
solution is the fitting formula for the nonlinear power spectrum by
\citet{pea96}, which is based on an idea originally proposed by
\citet{ham91}.  In the strongly nonlinear regime, on which we focus in
this paper, the Peacock-Dodds formula agrees well with the simulations
for $\Omega_0 = 1$, and also for $\Omega_0 < 1$ and $\Omega_0 +
\lambda_0 = 1$.  We have found, however, that in $\Omega_0 = 0.1$ and
$\lambda_0 = 0$ cases the formula fails to fit the simulations.  This
may be understood as follows.  In such cases the slope of $\xi$
asymptotically approaches that in conventional self-similar solution
only on scales where $\xi$ is extremely large.  This means that the gap
between these asymptotic scales and linear scales is rather big, and it
cannot be simply interpolated.  In contrast our model predicts the shape
of $\xi$ up to the scale corresponding to $\xi \sim$ a few hundred, so
interpolation between this scale and the linear scale may be much
easier.  Thus our model may be useful in improving the Peacock-Dodds
formula, especially in $\Omega_0 <1$ and $\lambda_0 = 0$ cases.

In summary, although our proposed model still needs to be improved, the
degree of success for $\xi \gtrsim 200$ is encouraging, and useful at
least empirically.  We hope that the idea of quasi-self-similar
evolution may give a useful insight towards a better understanding of
nonlinear dynamics of the mass density field in the universe. In
particular, we plan to improve the existing fitting formula for the
nonlinear power spectrum \citep{pea94,pea96} by applying the
quasi-self-similar ansatz.

\acknowledgments 

We are grateful to Y.~P.~Jing for kindly providing his results of N-body
simulations.  A.~T. acknowledges support from Research Fellowships of
the Japan Society for the Promotion of Science.  This research was
supported in part by the Grant-in-Aid by the Ministry of Education,
Science, Sports and Culture of Japan (07CE2002, 12640231) to RESCEU, and
by the Supercomputer Project (No.99-52, No.00-63) of KEK.

\clearpage

\begin{deluxetable}{cclcc}
 
 \tablecaption{Simulation parameters. \label{tab:param}} 
 \tablewidth{0pt}

 \tablehead{\colhead{$\Omega_0$} & \colhead{$\lambda_0$} & \colhead{$n$}
 & \colhead{$N \,{}^{\rm a}$} & \colhead{$\varepsilon_{\rm grav}/L_{\rm box} {}^{\rm b}$} }
 \startdata
   1.0 & 0.0 & $-1$, $-2$ \qquad & $256^3$ &
   $5.9 \times 10^{-4}$ \\
   0.1 & 0.0 & $-1$, $-2$ & $200^3$ &
   $7.5 \times 10^{-4}$ \\
   0.1 & 0.9 & $-1$, $-2$ & $200^3$ &
   $7.5 \times 10^{-4}$ \\
 \enddata
\bigskip
\renewcommand{\arraystretch}{0}
\begin{tabular}{r@{}p{16cm}}
   $^{\rm a}$ & The number of particles.\\
   $^{\rm b}$ & Gravitational softening length in units of the box
                size $L_{\rm box}$.\\ 
\end{tabular}
\end{deluxetable}

\clearpage

\end{document}